\documentclass{eas}
\usepackage{graphicx}
\begin{document}

\title{Status and Prospects of the DMTPC Directional Dark Matter Experiment}
\author{Jocelyn Monroe for the DMTPC Collaboration}\address{Department of Physics, Royal Holloway University of London}
\begin{abstract}
 The DMTPC directional dark matter detection experiment is a low-pressure CF4 gas time projection chamber, instrumented with charge and scintillation photon readout. This detector design strategy emphasizes reconstruction of WIMP-induced nuclear recoil tracks, in order to determine the direction of incident dark matter particles. Directional detection has the potential to make the definitive observation of dark matter using the unique angular signature of the dark matter wind, which is distinct from all known backgrounds. This talk will briefly review the experimental technique and current status of DMTPC.
\end{abstract}
\maketitle
\section{Introduction}

 Despite strong astrophysical evidence that dark matter comprises
 approximately 23$\%$ of the universe (Spergel \etal~\cite{Spergel:2006hy}), the
 nature of this dark matter remains largely unknown.  Weakly
 interacting massive particles (WIMPs) are a favored dark matter
 candidate.  Direct WIMP detection experiments search for the
 interaction of WIMPs with a nucleus in the detector, resulting in
 low-energy nuclear recoils.  Most experiments seek to detect the
 kinetic energy deposited by the recoiling nucleus; a handful of
 recent efforts, including this work, also seek to detect the
 direction of the nuclear recoil, and in this way, infer the direction
 of incoming WIMPs (Ahlen \etal~\cite{Ahlen:2009ev}).  

The arrival direction of WIMPs is predicted to peak in the direction
opposite to the earth's motion around the galactic center, and have a
time-varying asymmetry in the forward vs. backward direction
distribution because of the Earth's rotation
(Spergel~\cite{Spergel:1987kx}). No background is expected to have the
same angular distribution and modulation as a dark-matter induced
signal, and therefore directional detection offers the potential for
unambiguous observation of dark matter (Green
\etal~\cite{Green:2006cb}).  This paper discusses recent progress from
the Dark Matter Time Projection Chamber (DMTPC) collaboration toward a
directional detection experiment.

\section{DMTPC Experiment}

\begin{figure*}
\centering
\includegraphics[height=6.cm, angle=90]{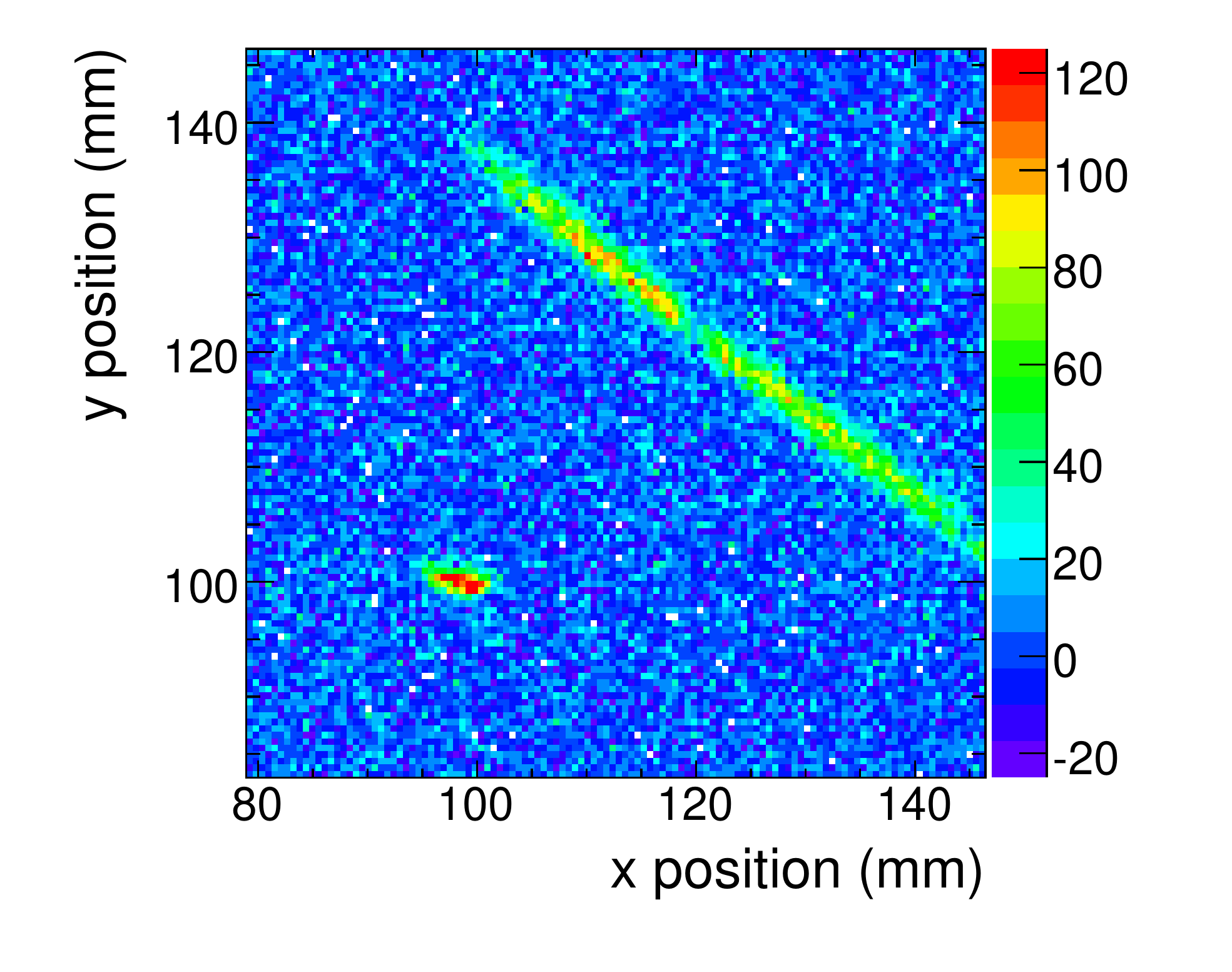}
\setlength{\abovecaptionskip}{0pt}
\caption{A CCD image showing a sub-region of a background-subtracted event from the top camera containing a segment of an alpha track (long) and a candidate nuclear recoil (short); intensity is in units of ADU, indicated by color. \label{fig:fig1}}
\setlength{\belowcaptionskip}{0pt}
\end{figure*}

DMTPC is a dark matter detector designed to measure the direction and
energy of recoiling nuclei.  The dark matter target is CF$_4$,
selected for the relatively large predicted axial-vector coupling for
fluorine, allowing sensitivity to spin-dependent WIMP
interactions (Lewin \& Smith~\cite{lewin1995}) with relatively small target masses.
The detector consists of two optically isolated back-to-back low-pressure time
projection chambers, with a 14.6$\times$14.6$\times$19.7~cm$^{3}$
(15.9$\times$15.9$\times$19.7~cm$^{3}$) fiducial volume for the top
(bottom) TPC.  The TPCs are filled with 75$\pm$0.1 Torr of CF$_4$ gas
corresponding to a 3.3~g (2.85~g) fiducial mass of CF$_4$ (F).  In
75~Torr of CF$_4$, a recoiling fluorine nucleus with 50~keV kinetic
energy travels approximately 1~mm before stopping.  The cathode and
ground planes of the TPC are meshes with 256~$\mu$m pitch.  The
grounded mesh sits 0.5~mm from the copper anode plate.  Ionization
electrons from interactions in the fiducial volume drift in a uniform
electric field towards the anode region where avalanche multiplication
amplifies the electron signal and produces scintillation.
The gas gain is approximately $4 \times 10^4$, measured with an
$^{55}$Fe calibration source.  The photon/electron production ratio in
the avalanche region is 0.34$\pm$0.04, and the wavelength spectrum of
the scintillation light peaks at $\sim$600 nm (Kaboth \etal~\cite{Kaboth:2008mi}).
Scintillation light produced in the amplification region is focussed
by a Nikon photographic lens (f/1.2) onto an Apogee Alta U6 camera
containing a 1024$\times$1024 element Kodak 1001E CCD chip.
To improve the signal-to-noise ratio,
pixels are binned 4$\times$4 prior to digitization.  Example alpha and nuclear recoil tracks are shown in Fig.~\ref{fig:fig1}.  In addition to optical readout, the charge induced on the ground mesh and anode electrodes is also digitized.  For a more detailed discussion of the 10 liter detector see (Ahlen \etal~\cite{Ahlen:2010ub}).  

The detector is calibrated with alpha and gamma sources.  Track length
and energy calibrations employ $^{241}$Am alpha sources at fixed
locations in the top and bottom TPCs.  
The length calibration relies on the known horizontal separation,
2.5$\pm$0.1 cm, of resistive separators in the amplification region.
Imaging the spacers gives the length calibration of
143$\pm$3~$\mu$m$\times$143$\pm$3~$\mu$m
(156$\pm$3~$\mu$m$\times$156$\pm$3~$\mu$m) for each CCD pixel of the
top (bottom) anode.
The energy calibration analysis compares the integral light yield of
segments of alpha tracks at known distances from the source, in the
arbitrary digital units (ADU) of the CCD, with the SRIM
simulation (Ziegler, Biersack, \& Littmark~\cite{srim}) prediction for the visible energy loss in that
segment.  The segment length is chosen such that the SRIM prediction
for the energy loss in each segment is 100-1000 keV, depending on the
location and size of the segment along the alpha track.  
According to SRIM, at these energies, the alpha energy loss is $>$97\% electronic and so we are not sensitive to assumptions about the nuclear quenching in this calibration.  This procedure gives the energy calibration of
9.5$\pm$0.5 and 12.9$\pm$0.7 ADU/keV respectively for the top and
bottom cameras.
The system gain (ADU/keV) may vary with position and in time.  The
gain non-uniformity across the field of view due to local variations
of the amplification is measured with a 14 $\mu$Ci $^{57}$Co source,
which provides uniform illumination of the field-of-view from
scattered 122 keV photons, integrated over 10,000 seconds.  The
measurement yields a 10$\%$ variation of the total system gain, which
is included as a position-dependent correction.  The stability of the
gain vs. time was measured to be 1$\%$ over 24 hours using an
$^{241}$Am alpha source.  To maintain this, the chamber is evacuated
to 10 mTorr and refilled with CF$_4$ every 24 hours.
\begin{figure*}
\centering
\includegraphics[height=6.cm, angle=90]{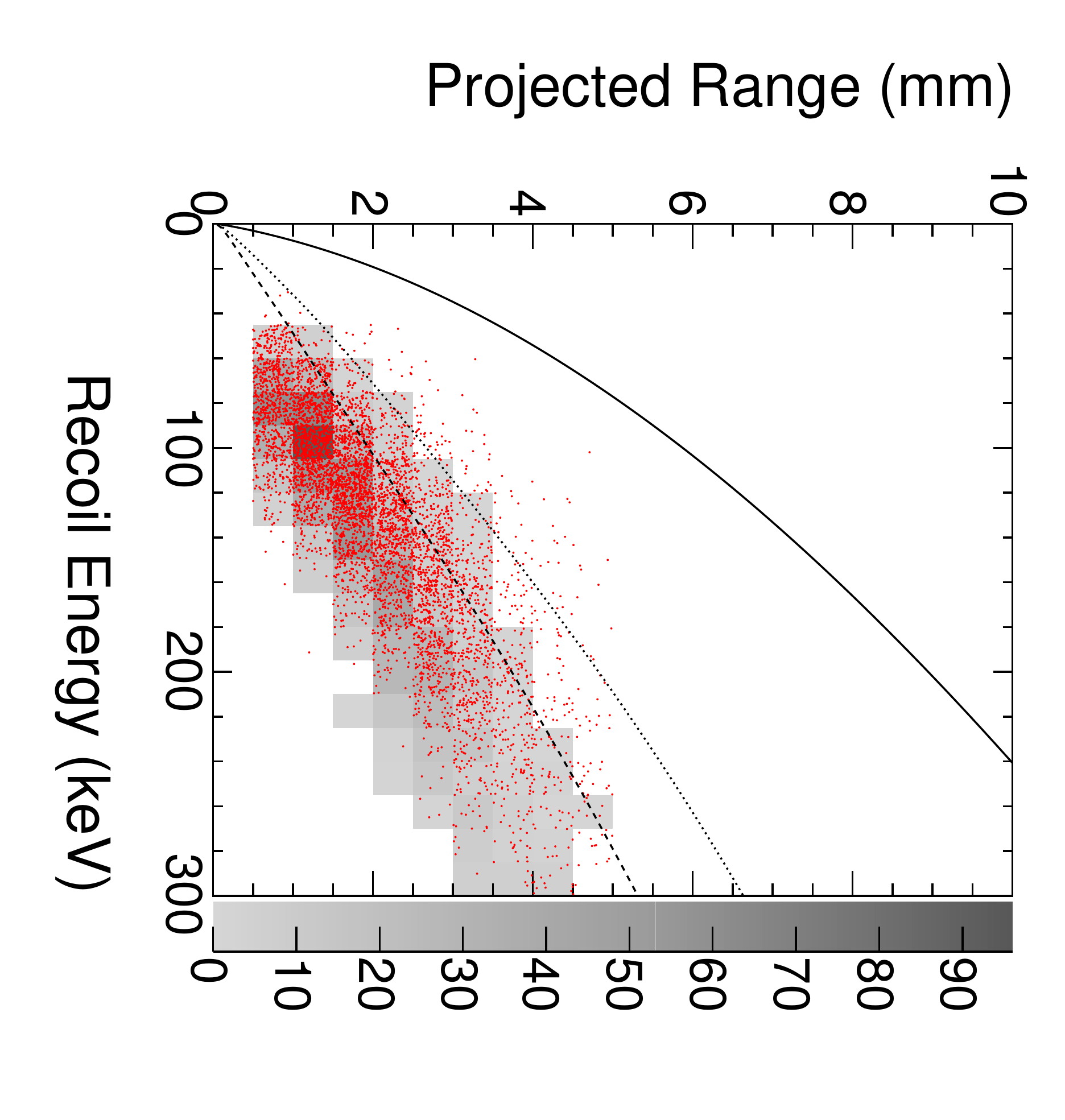}
\setlength{\abovecaptionskip}{0pt}
\caption{Reconstructed projected range (mm) vs. reconstructed energy (keV) for WIMP-search data (black points) compared with 200 GeV/c$^2$ WIMP-F Monte Carlo (shaded squares) after nuclear recoil selection cuts.  Lines are SRIM predictions for the maximum projected range vs. energy for helium (solid), carbon (dotted), and fluorine (dashed).  \label{fig:fig2}}
\setlength{\belowcaptionskip}{0pt}
\end{figure*}
\begin{figure*}
\centering
\includegraphics[height=6.cm, angle=90]{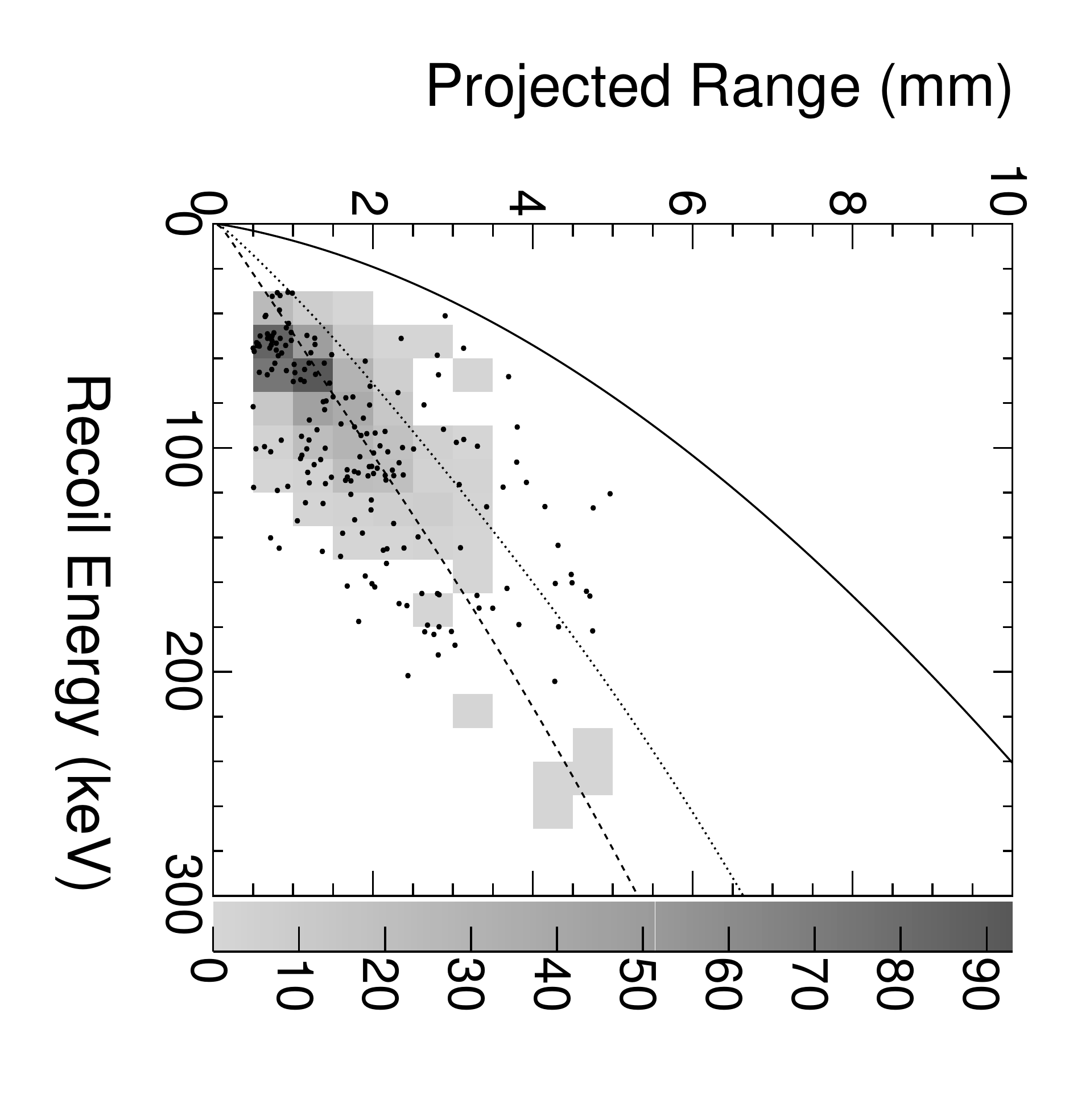}
\setlength{\abovecaptionskip}{0pt}
\caption{Reconstructed projected range (mm) vs. reconstructed energy (keV) for $^{252}$Cf calibration data (red points) compared with $^{252}$Cf-F Monte Carlo (shaded squares) after nuclear recoil selection cuts.  Lines are SRIM predictions for the maximum projected range vs. energy for helium (solid), carbon (dotted), and fluorine (dashed).  \label{fig:fig3}}
\setlength{\belowcaptionskip}{0pt}
\end{figure*}%

In the track reconstruction, raw CCD images are first background
subtracted using an average of 100 dark frames (collected every 1000
exposures).  Events which are consistent with sparking and CCD noise,
such as spark-induced RBI backgrounds, are removed (Rest \etal~\cite{rbis}).  The
track finding algorithm identifies groups of five contiguous bins with
at least $3.7\sigma$ counts per bin above the image mean as clusters;
3.7 is chosen to optimize the energy reconstruction.  The visible
energy is the integral of the counts in a track, divided by the energy
calibration constant (ADU/keV).  To convert visible to nuclear recoil
energy we use the CF$_4$ quenching factor calculated
in (Hitachi~\cite{Hitachi:2008kf}).  The projected range of a track on the
amplification plane is calculated from the maximally separated pixels
in the cluster, multiplied by the length calibration ($\mu$m/pixel).
The track angle projected on the amplification plane ($\phi$) is
determined by finding the major axis angle of an ellipse with the same
second moment as the pixels in the cluster.  The sense of the
direction is estimated from the skewness of the track light yield.
The recoil energy and angle reconstruction resolution are 15\% and
40$^o$ at 50 keVee (80 keV nuclear recoil energy).  The energy
resolution is measured with alpha calibration
data (Caldwell \etal~\cite{Caldwell:2009si}).  The angular resolution is estimated with
a Monte Carlo simulation of fluorine recoils from the $^{252}$Cf
calibration source.  The $^{252}$Cf-F Monte Carlo is compared with
data in Figure~\ref{fig:fig2}.  More detail on directionality studies
with this detector technology can be found in (Dujmic \etal~\cite{Dujmic:2008iq}).

To study detector backgrounds, a commissioning run was taken in a surface laboratory at MIT.  The background rates are summarized in Tab.~\ref{table:rates}.
We are pursuing rejecting backgrounds from interactions in the CCD
chip and spark-induced RBI by requiring coincidence of CCD and charge
readout signals, with consistent measured event energies between the
two independent channels.  Preliminary results indicate $10^{-2}$
rejection from the coincidence requirement (Lopez \etal~\cite{Lopez:2011yv}). In a
CCD-only analysis, these events are rejected in software, identified
as having large ADU and RMS of the pixels comprising the
track (Ahlen \etal~\cite{Ahlen:2010ub}).  
Background events inside the fiducial volume come primarily from
alphas and neutrons.  Alpha particles are emitted by radio-impurities
in or on the materials of the detector; the majority are from the
stainless steel drift cage.  These are rejected as CCD edge-crossing
tracks.  Background neutrons come primarily from $^{238}$U and
$^{232}$Th decays in and near the detector. The nuclear recoil
candidate events remaining after all background cuts are shown in
Fig.~\ref{fig:fig3}, compared to WIMP Monte Carlo.  These are
consistent with neutron-induced backgrounds (Ahlen \etal~\cite{Ahlen:2010ub}); for
comparison, Fig.~\ref{fig:fig2} shows calibration $^{252}$Cf
neutron-induced recoils.  There is no evidence for gamma-induced
electron backgrounds; the measured upper limit on the rejection is
$5.6 \times 10^{-6}$ between 40 and 200 keVee at 90\% confidence
level (Lopez \etal~\cite{Lopez:2011yv}).
\begin{table}
\centering
\begin{tabular}{cc}
\hline
\hline
Event Selection Cut & Rate (Hz) \\
\hline
All Tracks & 0.43\\
Residual Bulk Images & 0.15\\
CCD Interactions & 4.4$\times 10^{-3}$\\
Alpha Candidates & 8.2$\times 10^{-5}$\\
Nuclear Recoil Candidates in 80 $< E_R <$ 200 keV & 5.0$\times 10^{-5}$ \\
\hline
\hline
\end{tabular}
\caption{Surface run event rates (Hz) after each background rejection cut, using CCD information only, summed over cameras. 
\label{table:rates}}
\end{table}

Using the surface commissioning data we set a limit on the
spin-dependent WIMP-proton interaction cross section using the method
described in (Lewin \& Smith~\cite{lewin1995}).  The analysis energy window, 80-200
keV, is chosen to maximize the efficiency (estimated to be 70\% at 80
keV threshold energy).  There are 105 events passing all cuts in 80
$<$ $E_{recoil}$ $<$ 200 keV in a 35.7 gm-day exposure, with 74
predicted neutron background events based on the surface neutron
spectrum measurement in (Nakamura~\cite{neutrons}).  We set a 90\% confidence
level upper limit on the spin-dependent WIMP-proton cross section of
$2.0\times10^{-33}$ cm$^{2}$ at 115 GeV/c$^2$ WIMP mass, assuming zero
expected background.  Further details can be found
in (Ahlen \etal~\cite{Ahlen:2010ub}).

\section{Outlook and Conclusions}  
We have studied the backgrounds in the DMTPC 10 liter detector
prototype from a surface run, and present the first dark matter limit,
$\sigma_{\chi-p}$ $<2.0\times10^{-33}$ cm$^{2}$ at 90\% C.L.  DMTPC is
investigating the power of coincident charge and light readout to
further reject backgrounds. The 10L detector described here began
running underground at the Waste Isolation Pilot Plant outside
Carlsbad, NM in October 2010.  The depth of the WIPP site is 1.6 km
water-equivalent.  The gamma, muon, and radon levels have
been measured, and the neutron background has been estimated at this
site (Esch~\cite{Esch:thesis}).  Based on these, we project that underground
operation will lower the neutron background to $<$1
event/year.


\end{document}